\documentclass[aps,prl,twocolumn,groupedaddress,floatfix,showpacs]{revtex4-1}
\usepackage{amssymb,amsmath}
\usepackage{graphicx}
\usepackage{subfigure}
\usepackage[english]{babel}
\usepackage{float}
\usepackage{color}

\begin{document}

\title{Disorder-free weak dynamic localization in deformable lattices}

\author{Alexander V. Savin$^1$, Yuri S. Kivshar$^2$, and Mario I. Molina$^3$}

\affiliation{$^1$Semenov Institute of Chemical Physics, Russian Academy of Science, Moscow 117977, Russia\\
$^2$Nonlinear Physics Centre, Australian National University, Canberra ACT 2601, Australia\\
$^3$Departamento de F\'{\i}sica, Facultad de Ciencias, Universidad de Chile, Casilla 653, Santiago, Chile}

\begin{abstract}
We study the electron transport in a deformable lattice modeled in the semiclassical
approximation as a discrete nonlinear elastic chain where acoustic phonons are in thermal equilibrium
at temperature $T$. We reveal that an effective {\em dynamic disorder} induced in the system due
to thermalized phonons is not strong enough to produce Anderson localization. However,
for weak nonlinearity we observe a transition between ballistic (low $T$) and diffusive (high $T$) regimes,
while for strong nonlinearity the transition occurs between the localized soliton
(low $T$) and diffusive (high $T$) regimes. Thus, the electron-phonon interaction results
in weak temperature-dependent dynamic localization.
\end{abstract}

\maketitle

{\em Introduction.}
Propagation of an excitation across a deformable medium is of fundamental importance in condensed
matter physics, chemistry, and biology, and this topic is associated with a long history~\cite{ziman}.
In particular, the electron-phonon interaction lies at the heart of transport phenomena in solids
that determine conductivity of metals~\cite{grimvall1,grimvall2}, energy transfer along
protein molecules~\cite{davydov}, superconductivity~\cite{Schrieffer}, polaron transport~\cite{alexandrov},
and many-body physics in general~\cite{mahan}.

One of the important ingredients that plays a prominent role in conductivity of any system
is {\em disorder}. In general, a quenched disorder introduced into parameters of a Hamiltonian system
tends to impede the propagation of excitations by creating a random superposition of many
coherent waves due to multiple reflections. This effect is known as {\em Anderson Localization} (AL).
For one-dimensional systems, the presence of any amount of disorder destroys conductivity,
and it turns the system into an insulator
\cite{disorder1,disorder2,disorder3,disorder7,disorder5,disorder6,disorder4,disorder7b,disorder9}.

One might wonder whether the presence of quenched disorder is a requisite for localization.
Some systems free of disorder but experiencing many-body interactions can display a degree
of localization, if special inhomogeneous initial states are used~\cite{yao3,yao2,yao}.
As an example, we mention the models containing heavy and light interacting particles 
where the heavy particles provide an effective disorder for the light particles.
This coupling results in a transient subdiffusive
behavior and ergodicity at long times~\cite{MBL}.

A recent analysis of many-body localization problems suggests a possibility of having AL
without the presence of a quenched disorder. In this case, the system generates its own disorder
dynamically~\cite{smith}. A recent study of high-dimensional models with nonrandom
quasiperiodic potentials reveals an intermediate diffusive phase between the ballistic and
localized phases for the three-dimensional Andre-Aubry model~\cite{devakul}.

In this Letter, we provide a new twist to an old problem of the electron-phonon interaction and consider
an electron coupled to a deformable nonlinear chain modeled, in the semiclassical approximation, as
a discrete elastic 
\begin{figure}[t!]
\begin{center}
\includegraphics[angle=0, width=0.8\linewidth]{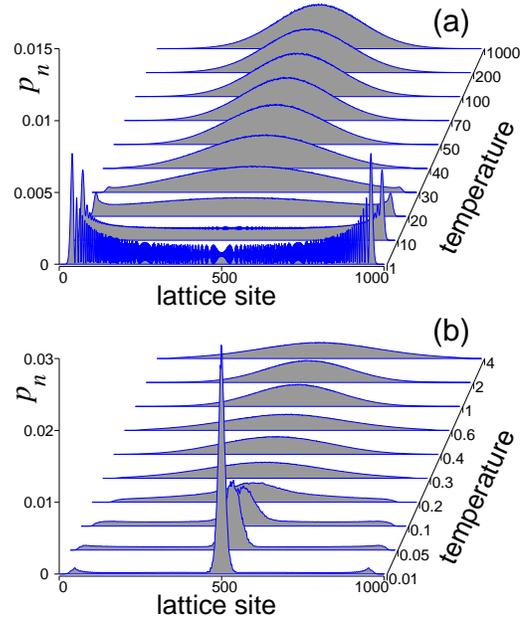}
\end{center}
\caption{\label{fg01}\protect
Electron distribution $p_n$ for (a) weak  and (b) strong electron-phonon interaction
at different temperatures. Ballistic-to-diffusive (a) and solitonic-to-diffusive (b) motion is observed  
(see the main text for data values).\label{fig1}
}
\end{figure}
array supporting acoustic phonons in thermal equilibrium.
We demonstrate the existence of a localization transition from a ballistic regime, at low temperatures,
to a diffusive regime, at high temperatures, and weak nonlinearity.
For strong nonlinearity, we observe the transition between localized (solitonic) and diffusive regimes (see Fig.\ref{fig1}). The transition is related to a resonance between two natural time scales present in the model.

{\em Model.} 
We consider an extra electron propagating along a deformable lattice
described by the Hamiltonian
\begin{equation}
H = H_{el} + H_{ph} + H_{int},
\label{f1}
\end{equation}
where the first term corresponds to the electron energy
$$
H_{el}=\sum_{n=1}^{N}\left[(\epsilon_{0}-2J)|\phi_{n}|^2
- J\phi_{n}^*(\phi_{n+1}+\phi_{n-1})\right],
$$
the second term represents the mechanical energy of the chain (the energy of a phonon sub-chain)
$$
H_{ph}=\sum_{n=1}^{N}\left[\frac12M \left({dx_{n}\over{d t}}\right)^{2}+V(\rho_{n})\right],
$$
and the third term corresponds to the interaction energy between
an electron and chain deformations
$$
H_{int} = -C\sum_{n=1}^{N}(\rho_{n}+\rho_{n-1})|\phi_{n}|^2.
$$
Here, $N$ is the number of particles (atoms) in the chain, $\{\phi_n\}_{n=1}^N$ --the electronic wave function in the lattice representation, normalized by the condition
\begin{equation}
\sum_{n=1}^N|\phi_n|^2=1,
\label{f2}
\end{equation}
$x_n$ is the displacement of the $n$th particle from its equilibrium position,
$\rho_n=x_{n+1}-x_n$ is the lengthening of the $n$th chain bond,
$\epsilon_{0}$ is the electron affinity to a chain particle,
$M$ is the particle mass, $J=\Delta E/4$ ($\Delta E$ --width of the conduction bond),
$C$ is the electron-phonon coupling. The molecular interaction potential is given by $V(\rho)=K\rho^2/2$,
where $K$ is the elasticity of the chain bond.

The equations of motion, corresponding to the Hamiltonian (\ref{f1}) have the form
\begin{eqnarray}
i\hbar {d\phi_n\over{d t}}&=&-J(\phi_{n+1}{+}\phi_{n-1})-C\phi_n(x_{n+1}{-}x_{n-1}),\nonumber
\\
M {d^2 x_{n}\over{d t^2}}&=&K(x_{n+1}{-}2x_n{+}x_{n-1})-C(|\phi_{n+1}|^2{-}|\phi_{n-1}|^2).\nonumber
\end{eqnarray}
Let us introduce the dimensionless time $\tau=2Jt/\hbar$,
whose unit corresponds to the minimum possible time of the electron passage of one
chain link, and dimensionless displacement $u_n=x_n\sqrt{K/J}$. Then, we obtain the equations of motion
in the dimensionless form:
\begin{eqnarray}
i {d\phi_n\over{d \tau}} &=& -\frac12(\phi_{n+1}+\phi_{n-1})-\chi\phi_n(u_{n+1}-u_{n-1}),\label{f3}\\
m {d^2 u_n\over{d\tau^2}} &=& u_{n+1}-2u_n+u_{n-1}-\chi(|\phi_{n+1}|^2-|\phi_{n-1}|^2),\ \ \ \ \ \ \label{f4}
\end{eqnarray}
with the dimensionless mass $m=M(2J/\hbar)^2 K^{-1}$ and electron-phonon coupling
$\chi=C (2JK)^{-1/2}$. The corresponding dimensionless Hamiltonian reads
\begin{eqnarray}
{\cal H}&=& \frac12\sum_{n=1}^N\ [\ -\phi_n^*(\phi_{n+1}+\phi_{n-1})+m \left({du_{n}\over{d\tau}}\right)^2 \nonumber\\
&  & +(u_{n+1}-u_n)^2 - 2\chi(u_{n+1}-u_{n-1})|\phi_n|^2\ \ ].\label{f5}
\end{eqnarray}

{\em Electron dynamics in a thermalized chain.} We model the dynamics of an electron in
a thermalized chain of atoms in a periodic array of $N=1000$ particles where the electron
is placed at the site $n_0=N/2$ (the middle of the chain), with the initial
wave function $\phi_n(0)\equiv\delta_{n,n_0}$. To find the thermalized state,
we employ the Langevin equations,
\begin{eqnarray}
m {d^2 u_n\over{d \tau^2}}=u_{n+1}-2u_n+u_{n-1}-\chi(\delta_{n+1,n_0}-\delta_{n-1,n_0})\nonumber\\
  -\gamma m {d u_n\over{d\tau}}+\xi_n,~n=1,...,N,~~~\label{f6}
\end{eqnarray}
where the friction coefficient $\gamma=1/\tau_r$ ($\tau_r$=100 --relaxation time), 
$\xi_n(\tau)$ is an external normally distributed random force,
which satisfies the conditions $\langle \xi_n(\tau)\rangle =0$,
$\langle \xi_n(\tau_1)\xi_l(\tau_2)\rangle=2m\gamma T\delta_{nl}\delta(\tau_1-\tau_2)$
($T$ is the dimensionless temperature of the Langevin thermostat).
On time $\tau_0=10\tau_r$, we have a thermalized state of the phonon sublattice
at temperature $T$: $u_n^0=u_n(\tau_0)$, $v_n^0=(d u_n/d\tau)(\tau_0)$.

We integrate numerically the system of equations Eqs.~(\ref{f3}) and (\ref{f4}) with the initial conditions
$\{\phi_n(0)=\delta_{n,n_0},~u_n(0)=u_n^0,~(d u_n/d\tau)(0)=v_n^0\}_{n=1}^N$.
Accuracy of our numerical procedure is controlled 
by monitoring the integrals of motion (\ref{f2}) and (\ref{f5}).

For dimensionless  system  (\ref{f3}), (\ref{f4}) we have only {\em three parameters}:
particle mass $m$, electron-phonon coupling, $\chi$ and temperature $T$.
We are interested in the electron transport along this deformable chain,
as a function of these parameters. To characterize the temporal spreading of the electron distribution,
we focus on two observables: the width of the electron distribution
$$
L(\tau) = 1/\sum_{n=1}^{N}|\phi_{n}(\tau)|^4
$$
(for a completely localized state $\phi_n=\delta_{n,n_0}$, the width is $L=1$,
while for an extended state $\phi_n\equiv 1/\sqrt{N}$, the width is $L= N$)
and the electron root mean square (RMS) displacement
$$
R(\tau) = \left[\sum_{n=1}^{N}(n-n_0)^2 |\phi_{n}(\tau)|^2 \right]^{1/2}.
$$
The asymptotic behavior of the RMS displacement is usually of the form
$\langle R^2\rangle \sim t^{\alpha}$,
where the exponent $\alpha=0,0.5,1,1.5,2$ denote localized, subdiffusive, diffusive, superdiffusive
and ballistic regimes, respectively. In particular, for diffusion,
$\langle R^{2}(t)\rangle \sim D t$ where $D$ is called the diffusion coefficient.
A value averaged over $10^4$ thermal states $\{u_n^0,v_n^0\}_{n=1}^N$ is used to compute the average
electron distribution $p_n(\tau)=\langle |\phi_{n}(\tau)|^2\rangle$, $\langle L(\tau) \rangle$
and $\langle R(\tau)\rangle$.

For definiteness, we take the particle mass as $m=0.31$, then the velocity of acoustic phonons
(sound speed) $v_{ph}=\sqrt{1/m}=1.796$ will exceed the highest velocity of the electron, $v_{el}=1$.
For comparison, we also consider the chain with $m=35$ when $v_{ph}=0.169<v_{el}$.
\begin{figure}[t]
\begin{center}
\includegraphics[angle=0, width=1\linewidth]{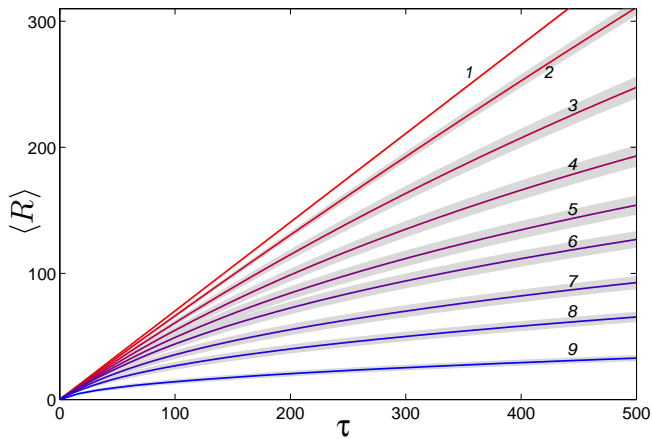}
\end{center}
\caption{\label{fg02}\protect
Time evolution of the RMS displacement $\langle R(t)\rangle$ in a chain for temperatures
 $T$=1, 10, 20, 30, 40, 50, 70, 100, and 200 (curves 1, 2, 3, 4, 5, 6, 7, 8, and 9).
Gray regions correspond to variability intervals
$[\langle R\rangle-\sigma,\langle R\rangle+\sigma]$, where
$\sigma^2=\langle R^2\rangle-\langle R\rangle^2$ (chain parameters $m=0.35$, $\chi=0.1$).
}
\end{figure}

Dynamics of the electron in a thermalized chain for different values of temperature $T$ is shown
in Fig.~\ref{fg01}. In the case (a), with weak electron-phonon interaction $\chi$=0.1, we have in increasing order (temperature $T$=1, time $\tau$=470);  (10, 500); (20, 520); (30, 540); (40, 560); (50, 600); (70, 1000); (100, 2000); (200, 8000); and (1000, 60000). In the case(b), with strong electron-phonon interaction $\chi$=1, we have in increasing order ($T$=0.01, $\tau$=480); (0.05, 500); (0.1, 500); (0.2, 500); (0.3, 550); (0.4, 600); (0.6, 900); (1, 2000); (2, 8000); and (4, 32000). For a chain with weak electron-phonon coupling ($\chi=0.1$) and low temperatures
($T=1$, 2), the electron wave function spreads with the maximum speed $v=1$,
signaling  a ballistic electron propagation with negligible effective coupling to thermalized phonons.  When the temperature grows, the dynamics changes, and for $T\ge40$
the wave function expands much slower. When the electron-phonon interaction becomes strong ($\chi=1$),
at low temperatures ($T\le 0.1$) we observe the formation of a {\em nonlinear self-localized state},
that expands slowly in the regime of large temperatures ($T\ge 0.5$).
\begin{figure}[t]
\begin{center}
\includegraphics[angle=0, width=1\linewidth]{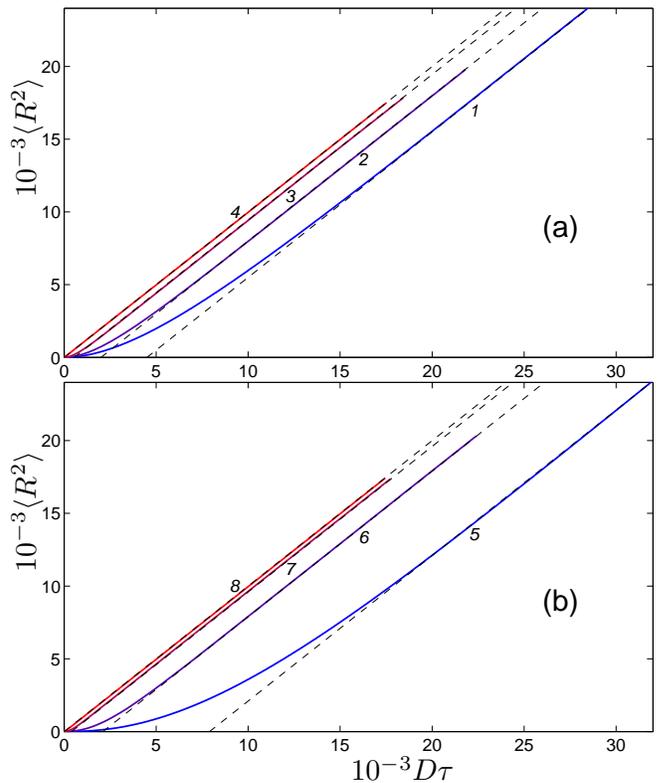}
\end{center}
\caption{\label{fg03}\protect
Time evolution of $\langle R(\tau)^2\rangle$ for a chain
with (a) weak electron-phonon interaction ($\chi=0.1$) and (b) strong
interaction ($\chi=1$). Curves 1, 2, 3, and 4 -- for temperatures
$T=40$ (diffusion coefficient $D$=56.6), 50 ($D$=36.4), 70 ($D$=18.4) and 100 ($D$=8.9).
Curves 5, 6, 7 and 8 -- for temperatures $T=0.4$ ($D$=54.5), 0.6 ($D$=24.9),
1 ($D$=8.9), and 2 ($D$=2.2).
}
\end{figure}

The study of the dependence of the RMS displacement vs. time,  $\langle R(\tau)\rangle$,
suggests that in a chain with weak electron-phonon interaction ($\chi=0.1$) and low temperatures,
a ballistic motion of the electron is observed: $\langle R(\tau)\rangle\sim\tau$ --see Fig.~\ref{fg02}.
However, for higher temperatures, this motion becomes diffusive, 
$\langle R(\tau)^2\rangle\sim D\tau$ --see Fig.~\ref{fg03}.
The critical value of temperature for the onset of normal diffusion 
depends on the electron-phonon coupling coefficient and particle's mass.
For $m=0.31$ and $\chi=0.1$, the normal diffusion occurs when $T\ge 40$,
for $m=35$ and $\chi=0.1$  --when $T\ge 2$, and for $m=0.31$ and $\chi=1$ --when $T\ge 0.4$.

\begin{figure}[t!]
\begin{center}
\includegraphics[angle=0, width=1\linewidth]{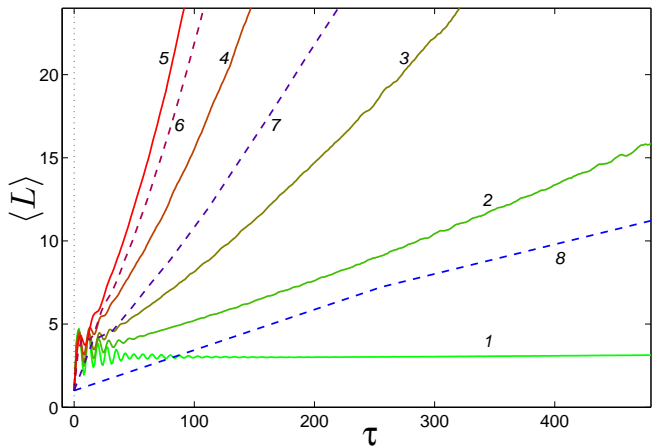}
\end{center}
\caption{\label{fg04}\protect
Time evolution of the average wavefunction width $\langle L(\tau)\rangle$
for a chain with strong
electron-phonon interaction $\chi=1$ for temperatures
 $T=0.002$, 0.05, 0.1, 0.2, 0.4, 0.6, 1, and 2 (curves 1, 2, 3, 4, 5, 6, 7, and 8).
}
\end{figure}

For strong electron-phonon interaction ($\chi=1$) and low temperatures, we observe the formation
of a self-localized state (``polaron''\ or ``electro-soliton''). To characterize better this case,
we study the temporal evolution of the width of the electron wave packet $\langle L(\tau)\rangle$.
As follows from Fig.~\ref{fg04}, for $T=0.002$ we observe the formation of a narrow localized soliton state
with a width $L\sim 3$, that slowly decays via expansion for larger values of temperature.
The highest mobility of the electron is achieved at $T=0.4$. For larger temperatures, we observe
the regime of {\em normal diffusion}, when the electron mobility decreases monotonically with temperature.
\begin{figure}[t]
\begin{center}
\includegraphics[angle=0, width=1\linewidth]{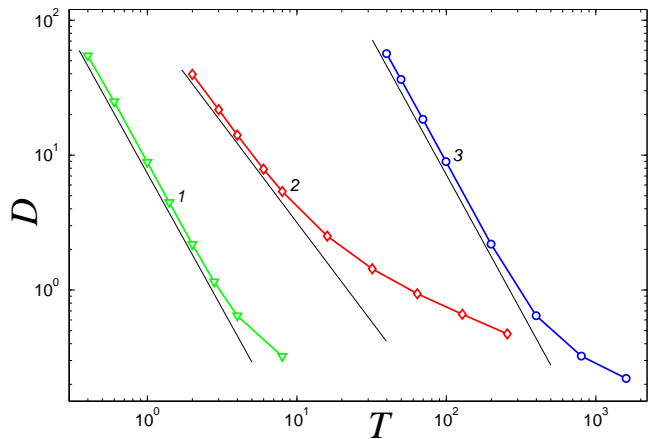}
\end{center}
\caption{\label{fg05}\protect
Dependence of the diffusion coefficient $D$ on temperature $T$ for chains 
with $m=0.35$, $\chi=1$ (curve 1); $m=35$, $\chi=0.1$ (curve 2), and
$m=0.35$, $\chi=0.1$ (curve 3). Straight lines correspond to the dependencies
$D=d/T^\alpha$ with $d=8.6$, $\alpha=2.00$; $d=110$, $\alpha=1.47$ and $d=98000$, $\alpha=2.02$,
respectively.
}
\end{figure}

Our numerical study of the thermalized chain indicates that there always exists a critical value
of temperature above which the electron dynamics is replaced by its diffusion characterized by
the diffusion coefficient $D=\lim_{\tau\rightarrow\infty}\langle R^2(\tau)\rangle/\tau$.
When $T$ grows, the diffusion coefficient $D$ decreases monotonously, so that
$D(T)\rightarrow 0$ for $T\rightarrow\infty$. As follows from Fig.~\ref{fg05},
the initial stage of this decay is characterized the power-like dependence
$D\sim d/T^\alpha$, where $\alpha>1$.

\begin{figure}[t!]
\begin{center}
\includegraphics[angle=0, width=1\linewidth]{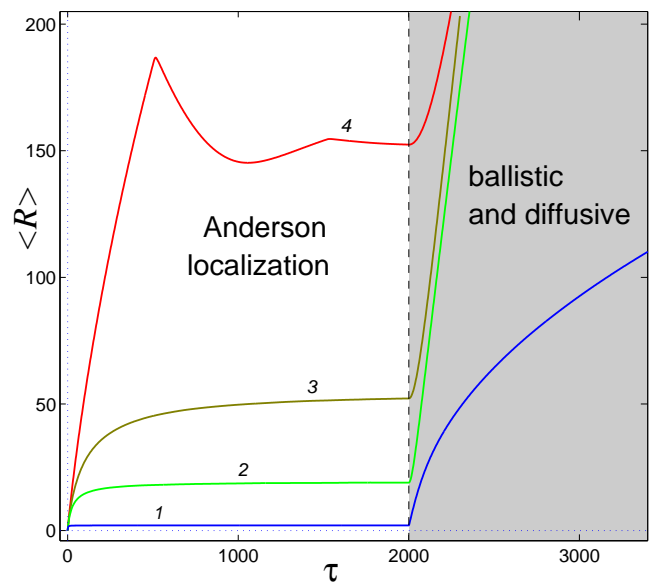}
\end{center}
\caption{\label{fg06}\protect
Time evolution of  $\langle R(\tau)\rangle$ for a chain with
temperature $T$=1, 0.1, 0.05 and 0.01 (curves 1, 2, 3, and 4).
For time $\tau<\tau_1=2000$ fixed particles positions $u_n(\tau)\equiv u_n^0$ in the 
system of equations (\ref{f3}). In this one-component regime, we have AL.
For time $\tau\ge\tau_1$, we use the two-component system 
(\ref{f3}), (\ref{f4}) with the initial conditions
$\{\phi_{n}(\tau_1),~u_{n}(\tau_1)=u_{n}^{0},(du_{n}/d\tau)(\tau_1)=v_{n}^{0}\}_{n=1}^{N}$.
In this, regime we have destruction of the Anderson localization. Chain parameters are $m$=0.35, $\chi$= 1.
}
\end{figure}

{\em Dynamic vs static localization}. For large values of temperature $T$, thermal fluctuations create a {\em non-stationary random potential}.
If we fix this effective potential at a certain time and consider ``frozen''\ fluctuations at time 
$\tau_0$, the position of particles will become fixed (we put $u_n(\tau)\equiv u_n(\tau_0)$).
Then, the problem is described by the one-component equation (\ref{f3}) with random diagonal coefficients.
This model with a diagonal disorder supports true {\em Anderson localization}, when the averaged
displacement of the electron tends to a finite value. Our numerical simulations confirm this
prediction since $\langle R(\tau)\rangle\rightarrow \mbox{const} <\infty$ for $\tau\rightarrow\infty$,
as shown in the left part of Fig.~\ref{fg06}.

Now, if at some time $\tau_{1}$,  we release the constraint on the particle's coordinates, the evolution equations will correspond to the two-component model (\ref{f3}) and (\ref{f4}), instead of one-component model (\ref{f3}), and  the stationary disorder will be replaced by dynamical disorder due to thermal fluctuations. This dynamical disorder leads to a decay of the true AL, as shown in the right part of Fig.~\ref{fg06}.
Therefore, in the case of dynamical disorder, the AL effects are replaced by ballistic or diffusive motion. The diffusion
speed is lower for higher values of temperature and high values of dynamic disorder,
and the electron becomes localized only in the limit $T\rightarrow\infty$.

{\em Discussion.}
The main result of this work is that the dynamics of the coupled electron-phonon system does
not give rise to a form of disorder strong enough to effect the Anderson localization.
Instead, we have a weak form of localization that manifest itself in the form of a ballistic
electronic transport at low temperatures or a diffusive regime at higher temperatures.
The effect of nonlinearity (electron-phonon coupling) produces a transition from ballistic
to diffusive regime  at weak nonlinearity, while at strong nonlinearity, it gives rise
to a transition from a solitonic regime to a diffusive regime.
All of these effects can be explained as follows. There is no Anderson localization for
our system in the general case (no frozen coordinates) for the following reasons:
For an acoustic chain, there exists  a number of phononic states with wavelengths 
comparable to the size of the system. These states are insensitive to disorder.
For these states, the relative displacement between two consecutive masses is nearly constant,
meaning that the effective equation of motion for the electron becomes
\begin{equation}
i \frac{d}{dt}\phi_{n}(t)\approx -\frac12(\phi_{n+1}+\phi_{n-1}) - B\phi_{n},
\label{f7}
\end{equation}
with $B$ nearly constant. This implies no dynamic Anderson localization, 
and ballistic transport. The effect is stronger at low $T$, since in these conditions
there is a substantial population of the long-wavelength modes. Now, at high $T$
the number of these states decreases and the transport becomes diffusive since the electron
undergoes a random walk transport. The second reason has to do with nonlinearity:
For small electron-phonon coupling, one  expects that nonlinearity effects
will weaken the superposition principle, because it inhibits the random coherent wave superposition.
This will have the effect of weaken any tendency towards Anderson localization.
At high nonlinearity on the other hand, we have soliton effects, not Anderson localization.
There will be partial self-trapping plus a component that propagates ballistically
(at low temperatures), or diffusively (at high temperatures).

In conclusion, we have studied the dynamic localization in a deformable lattice describing a motion of an electron in a discrete nonlinear elastic chain
with thermalized acoustic phonons. We have observed that this system does not support
Anderson localization due to {\em dynamic disorder}, but it demonstrates a transition between
ballistic and diffusive regimes, for small nonlinearity,
and the transition between the localized soliton and diffusive  regimes,
for strong nonlinearity.
Thus, the electron-phonon interaction results in weak disorder-free temperature-dependent
localization effects.

\acknowledgments
This work was supported by the Australian National University and Fondecyt Grant 1160177.

\end{document}